\documentclass[aps,pra,twocolumn,noshowpacs]{revtex4}
\usepackage{bm,graphicx,amsmath}
\begin{document}
\title{Alternative Experimental Protocol for a PBR-Like Result}
\author{D. J. Miller}
\email[]{d.miller@sydney.edu.au}
\affiliation{Centre for Time, University of Sydney, Sydney NSW 2006, Australia and School of Physics, University of New South Wales, Sydney NSW 2052, Australia}

\begin{abstract}
Pusey, Barrett and Rudolph (PBR) have recently proven an important new theorem in the foundations of quantum mechanics. Here we propose alternative experimental protocols which lead to the PBR result for a special case and a weaker PBR-like result generally. Alternative experimental protocols support the assumption of measurement independence required for the PBR theorem. 
\end{abstract}

\maketitle

\section{Introduction \label{intro}}
Pusey, Barrett and Rudolph (PBR) \cite{pbr} have proven a new theorem which imposes a significant constraint  on the interpretation of quantum mechanics \cite{nature}. The theorem shows that if one assumes that quantum systems possess physical states $\lambda$ which are solely responsible for determining the outcomes of experiments performed on the quantum systems, then no two distinct quantum states $|u \rangle$ and $|v \rangle$ can share the same physical state $\lambda$. It follows that any quantum state $|u \rangle$ is uniquely co-ordinated with a subset of the $\lambda$ and therefore should itself be regarded as a physical property of the quantum system. In the terms of Ref.~[\onlinecite{harrigan}], models of quantum mechanics that assume there are physical states are classified as $\psi$\textit{-epistemic} if there are two states which share the same physical state $\lambda$ and as $\psi$\textit{-ontic} if there is no pair of states which share the same $\lambda$. Thus PBR show that all such models of quantum mechanics are $\psi$-ontic.

Hall has shown that the assumptions used in the PBR theorem can be weakened and so the theorem can be generalised \cite{hall}. Colbeck and Renner \cite{colbeck} have reached the same conclusion as PBR via a different route and furthermore concluded that the quantum state is the \textit{only} physical property of the quantum system.

The experimental protocol for the PBR theorem \cite{pbr} requires a minimum of $N=2$ sources of the two arbitrary states $ |u \rangle$ and $|v \rangle$  that are involved. When $|\langle u |v \rangle| ^2>1/2 $, $N>2$ sources are required and as $|\langle u |v \rangle|\rightarrow 1 $, $N \rightarrow \infty$. As $N$ becomes larger, the experimental error that can be tolerated becomes smaller \cite{pbr}. As well as gates which operate on the individual states, the protocol requires an $N$-bit entangling gate operating without post-selection. It is interesting to investigate alternative experimental protocols to implement the PBR strategy for three reasons.

Firstly, alternative protocols may be easier to implement experimentally. Secondly, the PBR result is a very strong result because it applies to \textit{any} two quantum states. Proving the result for any two quantum states is necessary to deal with the formal definition of $\psi$-epistemic \cite{harrigan} (see above). However a $\psi$-epistemic model which literally had just two states which share the same $\lambda$ would be artificial and unphysical. Therefore $\psi$-epistemic defined in those terms imposes an unnecessarily harsh burden on experiments designed to distinguish between $\psi$-ontic and $\psi$-epistemic models. It has already been noted \cite{lewis} that one could require that a model is $\psi$-epistemic only if \textit{every} pair of non-orthogonal states share some physical states $\lambda$ (although that definition is not adopted in Ref.~[\onlinecite{lewis}]). It seems that it should be persuasive to prove that \textit{a significant proportion} of non-artificially chosen pairs of states do not share the same $\lambda$ because that would rule out all but artificially contrived $\psi$-epistemic models. Therefore it is worthwhile investigating protocols which lead to weaker versions of the PBR theorem. Thirdly, Hall \cite{hall} has pointed out that (i) the PBR theorem requires the assumption of measurement independence (MI), namely that the properties described by $\lambda$ are uncorrelated with the choice of the measurement performed on the states but (ii) only one measurement procedure for each pair of states is required so a limited form of the PBR theorem, not requiring MI but restricted to that measurement being performed, is possible \cite{hall}. Having a second experimental protocol available supports the assumption that the PBR theorem applies independently of the fate of the quantum systems involved.
 
\section{Alternative experimental protocols}

\subsection{XYZ Heisenberg Hamiltonian}

The general approach of this Section depends entirely on the strategy set out in Ref. \onlinecite{pbr}. We assume that there is a physical state $\lambda$ associated with a quantum system and that $\lambda$ alone is responsible for the outcome of any experiment on the quantum system. The quantum state $|u \rangle$ associated with the quantum system allows the probabilities of the experimental outcomes to be determined. If a quantum system can possess the same physical state when it is in the state $|u \rangle$ as when it is in the state $|v \rangle$, the two states $|u \rangle$ and $|v \rangle$ will be said to be conjoint.  The set of physical states that $|u \rangle$ and $|v \rangle$ could have in common will be called $\{\lambda\}_{uv}$. States which have no physical states in common, i.e. $\{\lambda\}_{uv}$ is empty, will be said to be disjoint. The set $\{\lambda\}_{uv}$ must be empty for orthogonal states because, for certain measurements, they never produce the same result, i.e. orthogonal states are disjoint and so there is no need to consider orthogonal states further here. 

Two arbitrary, distinct states $|u \rangle$ and $|v \rangle$ lie in a two-dimensional subspace and can be mapped onto spin states on a Bloch sphere. If
\begin{equation}
\langle u |v \rangle = \cos \theta e^{i\phi} ,
\end{equation}
so that
\begin{equation}
\label{theta}
\theta = \cos^{-1} |\langle u | v \rangle| \text{  and  } \phi = \arg \langle u |v \rangle 
\end{equation}
we can write the states, in the laboratory co-ordinate system, as
\begin{subequations}
\begin{align}
|u \rangle & =e^{-i\phi}(\cos \theta/2 |+ \rangle - \sin \theta/2  |- \rangle) \\
|v \rangle & =\cos \theta/2 |+ \rangle + \sin \theta/2 |- \rangle
\end{align}
\end{subequations}
where $|+ \rangle$ ($|- \rangle$) is spin up (down) along the $z$-axis (in units of $\hbar/2$) and $0<\theta<\pi/2$ because we consider distinct, non-orthogonal states. We will also be concerned with the state $|\overline{v} \rangle$ which is orthogonal to $|v \rangle$ in the subspace spanned by $|u \rangle$ and $|v \rangle$,
\begin{equation}
|\overline{v} \rangle= - \sin \theta/2 |+ \rangle + \cos \theta/2 |- \rangle .
\end{equation}
The states are shown in Fig.~1(a).
\begin{figure}
\includegraphics[scale=0.6]{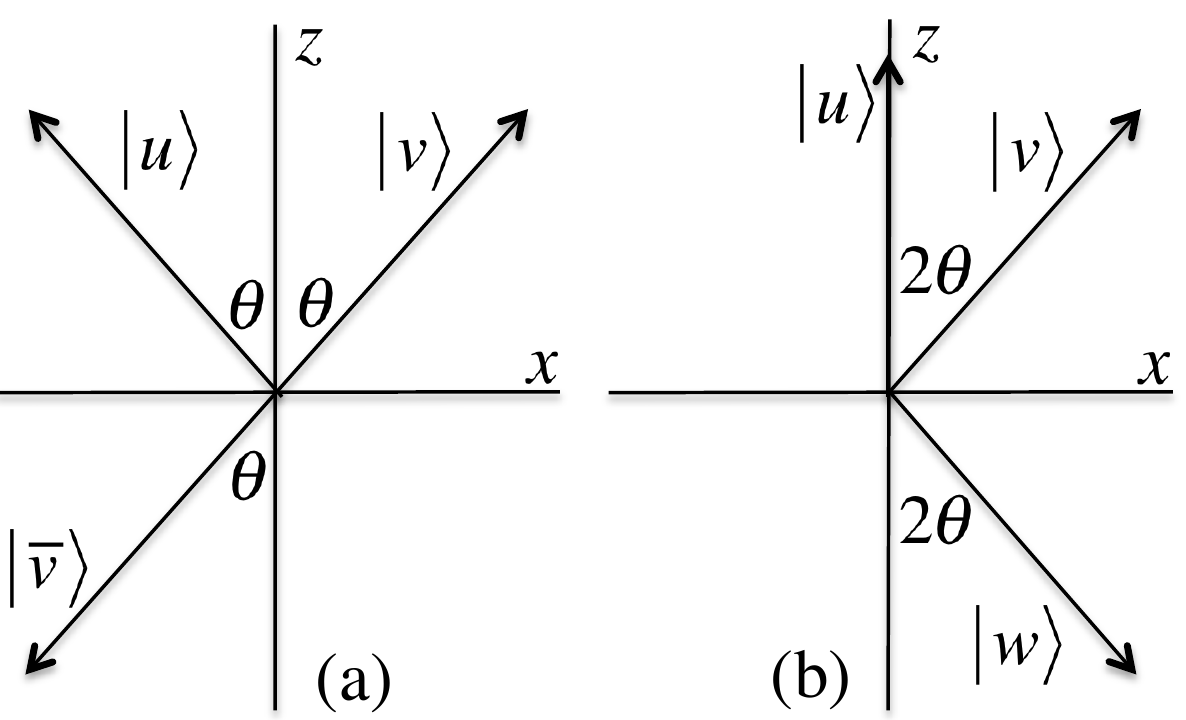}
\caption{The spin states involved in the present experimental protocols represented on a Bloch sphere referred to the laboratory co-ordinate system. (a) XYZ Heisenberg Hamiltonian. (b) Interaction with spin-orbit coupling.}
\end{figure}

In each run of the experiment, Alice produces either the state $|u \rangle_A$ or $|v \rangle_A$ and Bob produces either the state $|u \rangle_B$ or $|\overline{v} \rangle_B$. The spins from Alice and Bob are allowed to interact via a Heisenberg XYZ Hamiltonian
\begin{equation}
\label{H}
H=a\sigma^A_x \sigma^B_x+ b\sigma^A_y \sigma^B_y + c\sigma^A_z \sigma^B_z
\end{equation}
where, with energy measured in appropriate units, $a \neq \pm b$ and $c$ are real, dimensionless constants and  $\sigma^A_i$ and $\sigma^B_i$ are the Pauli matrices in the laboratory co-ordinate system. The eigenvalues $E_i$ and eigenstates $| {{e_i}} \rangle$ of $H$ are 
\begin{align}
E_1  & =  a - b + c, & | e_1 \rangle  = \frac{1}{\sqrt 2 }( |  +  \rangle _A| +  \rangle _B + |  -  \rangle _A| -  \rangle _B) \nonumber \\
E_2  & =   -a + b + c, & | e_2 \rangle  = \frac{1}{\sqrt 2 }( |  +  \rangle _A| +  \rangle _B - |  -  \rangle _A| -  \rangle _B) \nonumber \\
E_3  & =  a + b - c, & | e_3 \rangle  = \frac{1}{\sqrt 2 }( |  +  \rangle _A| -  \rangle _B + |  -  \rangle _A| +  \rangle _B) \nonumber \\
E_4  & =   -(a + b + c), & | e_4 \rangle  = \frac{1}{\sqrt 2 }( |  +  \rangle _A| -  \rangle _B - |  -  \rangle _A| +  \rangle _B) 
\end{align}
The eigenstates are the Bell states and they are non-degenerate because $a \neq \pm b$. In each run, the experiment consists of determining which of the eigenstates $| e_i \rangle$ is occupied. This could be done by switching the interaction in Eq.~(\ref{H}) on for a period of time so that each of the four joint states, $|u \rangle_A|u \rangle_B$, etc, evolves into a combination of the Bell states that it is not orthogonal with, removing the interaction and performing a Bell measurement. Each of the four joint states that can be produced by Alice and Bob are orthogonal to one of the $| e_i \rangle$:
\begin{align}
\label{eorthog}
\langle e_4 |u\rangle _A|u \rangle_B =0, \text{      } & \langle e_2 |u\rangle _A |\overline{v} \rangle_B =0,  \nonumber \\
\langle e_3 |v\rangle _A|u \rangle_B =0, \text{      } & \langle e_1 |v\rangle _A|\overline{v} \rangle_B =0 .
\end{align}

Assuming Alice and Bob produce their states independently (this assumption can be weakened \cite{hall}), the joint physical state $\lambda_{AB} = \lambda_A \lambda_B$. Now assume that $|u \rangle$ and $|v \rangle$ are conjoint and that $|u \rangle$ and $|\overline{v} \rangle$ are conjoint. Consider a run of the experiment in which $\lambda_A \in \{\lambda\}_{uv}$ and $\lambda_B \in \{\lambda\}_{u\overline{v}}$.  It is not possible for  $\lambda_{AB}$ to produce any of the outcomes because (i) $\lambda_{AB}$ is produced in conjunction with one of the four joint states $|u \rangle_A|u \rangle_B$, etc, (ii) $\lambda_{AB}$ alone produces the outcome, (iii) according to Eq.~(\ref{eorthog}), each of the four joint states forbids one of the outcomes and (iv) sooner or later $\lambda_{AB}$ will produce one of the outcomes which is forbidden by the state it is produced with.  But one of the outcomes will occur, i.e. one of the eigenstates will be occupied, in any run of the experiment. Therefore one of the assumptions must be false. We can conclude that, for any two states $|u \rangle$ and $|v \rangle$, either $|u \rangle$ and $|v \rangle$ are disjoint or $|u \rangle$ and $|\overline{v} \rangle$ are disjoint, or both pairs are disjoint. In other words, if $|u \rangle$ is conjoint with $|v \rangle$ then it is disjoint with the state orthogonal to $|v \rangle$ in the subspace spanned by $|u \rangle$ and $|v \rangle$.

\subsection{Interaction with spin-orbit coupling}

We next consider spins interacting via the Hamiltonian
\begin{equation}
H'=a\sigma^A_x \sigma^B_x+ b\sigma^A_y \sigma^B_y + c\sigma^A_z \sigma^B_z + d(\sigma^A_x \sigma^B_z - \sigma^A_z \sigma^B_x)
\end{equation}
where, with energy measured in appropriate units, $a$, $b$, $c$ and $d$ are real, dimensionless constants. The eigenvalues $E'_i$ and eigenstates $| {{e'_i}} \rangle$ of $H'$ are
\begin{widetext}
\begin{eqnarray}
E'_1  & = &  -a + b + c, \quad \; | e'_1 \rangle  = \frac{1}{\sqrt 2 }( |  +  \rangle _A| +  \rangle _B - |  -  \rangle _A| -  \rangle _B) \nonumber \\
E'_2  & = &  a + b - c, \quad \quad  | e'_2 \rangle  = \frac{1}{\sqrt 2 }( |  +  \rangle _A| -  \rangle _B + |  -  \rangle _A| +  \rangle _B) \nonumber \\
E'_3  & = &  -b - \sqrt{(a+c)^2 + 4d^2}, \nonumber  \\
& & | e'_3 \rangle  = \frac{1}{\sqrt 2 }[ \cos \alpha (|  +  \rangle _A| +  \rangle _B + |  - \rangle _A| -  \rangle _B)  + \sin \alpha ( |  +  \rangle _A| -  \rangle _B - |  -  \rangle _A| +  \rangle _B) ] \nonumber \\
E'_4  & = &  -b + \sqrt{(a+c)^2 + 4d^2}, \nonumber  \\
& & | e'_4 \rangle  = \frac{1}{\sqrt 2 }[ -\sin \alpha(|  +  \rangle _A| +  \rangle _B + |  - \rangle _A| -  \rangle _B) +  \cos \alpha( |  +  \rangle _A| -  \rangle _B - |  -  \rangle _A| +  \rangle _B) ]  
\end{eqnarray}
\end{widetext}
where
\begin{equation}
\label{alpha}
\tan \alpha = \frac{a + c + \sqrt{(a+c)^2 + 4d^2}}{2d}.
\end{equation}
The eigenstates are non-degenerate provided $a \neq c$.

This time we map the states $|u \rangle$ and $|v \rangle$ of Sect.~IIA onto the spin states
\begin{subequations}
\begin{align}
|u \rangle & =e^{-i\phi} |+ \rangle \\
|v \rangle & =\cos \theta |+ \rangle + \sin \theta |- \rangle
\end{align}
\end{subequations}
and also consider the state
\begin{equation}
|w \rangle =\sin \theta |+ \rangle + \cos \theta |- \rangle
\end{equation}
where again $0< \theta <\pi/2$.
The states are shown in Fig.~1(b). In each run of the experiment, Alice produces either the state $|u \rangle_A$ or $|v \rangle_A$ and Bob produces either the state $|u \rangle_B$ or $|w \rangle_B$. The spins from Alice and Bob are allowed to interact via $H'$ and a measurement is performed to determine which of the eigenstates $| e'_i \rangle$ is occupied. If the strengths of the components of spin-spin interaction can be chosen so that $\cos(\alpha + \theta)=0$, where $\alpha$ is given in Eq.~(\ref{alpha}) and $\theta$ in Eq.~(\ref{theta}), each of the four joint states that can be produced by Alice and Bob are orthogonal to one of the $| e'_i \rangle$:
\begin{align}
\langle e'_2 |u\rangle _A|u \rangle_B =0, \text{      } & \langle e'_4 |u\rangle _A |w \rangle_B =0,  \nonumber \\
\langle e'_3 |v\rangle _A|u \rangle_B =0, \text{      } & \langle e'_1 |v\rangle _A|w \rangle_B =0 .
\end{align}
Therefore by the same argument \cite{pbr} as in Sect.~IIA, we can conclude that if $|u \rangle$ is conjoint with $|v \rangle$ then it is disjoint with $|w \rangle$. When $\cos \theta=1/\sqrt{2}$, i.e. from Eq.~\ref{theta} when $|\langle u |v \rangle|^2 = 1/2$, $|v \rangle$ and $|w \rangle$ are the same state and so we can conclude that any pair of distinct states $|u \rangle$ and $|v \rangle$ are disjoint if $|\langle u |v \rangle|^2 = 1/2$. Thus the present protocol yields the same result as PBR for this special case.

\section{Discussion}

\subsection{Experimental implementation}

The Hamiltonians in Sec.~II are possible in principle because the spin-spin interaction tensor can take the required forms in $C_{2v}$ and most lower symmetries \cite{buck,robert}. Hamiltonians of the type proposed here have been considered in the quantum information processing (QIP) context.

The original proposals for QIP, e.g. quantum dots \cite{lossdi,burkardloss} and nuclear spins \cite{kane}, relied on the isotropic Heisenberg spin-exchange interaction. Sources of anisotropic contributions include spin-orbit coupling and the direct dipole-dipole interaction \cite{loss2,step}. They can be regarded as something to be corrected for \cite{loss2,step,bonesteel} or as an alternative means of creating entanglement with certain advantages \cite{lidar2,lidar3,lidar4}. Anisotropic exchange interactions of the type XY and XXZ arise in some systems proposed for QIP \cite{lidar1}. The Hamiltonian $H$ in Sec.~IIA is of the Heisenberg XYZ type which has been considered for QIP purposes more recently \cite{lidar4,abliz,kheir}. A Hamiltonian due to a pure dipole-dipole interaction can lead to a Hamiltonian of the XYZ form for a limited range of values of $a$, $b$ and $c$. The Hamiltonian $H'$ in Sec.~IIB can arise from an isotropic exchange interaction and the anti-symmetric part of the spin-orbit interaction, the Dzyaloshinskii-Moroya interaction, which is proportional to $\hat{\bm{n}}.[\bm{\sigma}^A \times \bm{\sigma}^B]$, with the unit vector $\hat{\bm{n}}$ joining the two spins in the $y$-direction. The Dzyaloshinskii-Moroya interaction is involved in systems proposed for QIP \cite{kavokin,chutia}. 

\subsection{Measurement independence}

Hall \cite{hall} has noted out that only one measurement procedure is involved in the PBR theorem. This is a significant difference compared with other ``no-go" theorems \cite{mermin} like the Bell inequalities and the Kochen-Spekker theorem which require at least two independently chosen experiments to be carried out. That means that the other ``no-go" theorems cannot be proven without the assumption of MI (i.e., that the physical state $\lambda$ is uncorrelated with the choice of the measurement(s) involved in proving the theorem). In the case of the PBR theorem, a limited form of the theorem applies without MI \cite{hall}, namely that the PBR theorem applies when the states involved are subject to the measurement procedure required to prove the theorem. 

The present results are relevant to MI because a PBR-like result can be obtained from the quite different measurement procedure proposed here. For the case $|\langle u |v \rangle|^2 = 1/2$, the PBR result follows from both the experimental procedure in Ref.~[\onlinecite{pbr}] and the experimental procedure in Sec.~IIB. The fact that the same conclusion can be drawn from different experimental procedures supports the idea that the conclusion applies independently of the measurement procedure, or indeed that the conclusion applies when no joint measurement of the quantum systems takes place. Of course, even if the fate of the quantum systems does not affect the PBR theorem that does not mean that MI applies to the other ``no-go" theorems. Hall has shown the other ``no-go" theorems are particularly vulnerable to even partially relaxing MI \cite{hall2}.  
 
\section{Conclusion} 

It has been shown that PBR-like results can be obtained with experimental protocols different from the one used in Ref.~[\onlinecite{pbr}] and that the PBR result itself can be obtained in the case when the overlap between the two states involved is $|\langle u |v \rangle|^2 = 1/2$. An advantage of the present experimental protocols is that only two sources of the states is required for all values of $|\langle u |v \rangle|^2$. From both protocols used here, one can conclude that if two states $|u \rangle$ and $|v \rangle$ are conjoint then $|u \rangle$ is disjoint with two related states $|\overline{v} \rangle$ and $|w \rangle$. The weaker result (except for the case $|\langle u |v \rangle|^2 = 1/2$) obtained here is not necessarily a disadvantage because it would be unreasonable to maintain the $\psi$-epistemic interpretation for some pairs of states when it is ruled out for most pairs of states. Finally, the existence of alternative experimental protocols leading to similar results to PBR, and the same result for a special case, supports the assumption of MI required for the PBR theorem.  

\section{Acknowledgements}

I am grateful to M. J. W. Hall for helpful comments.

\end{document}